# VLSI Implementation of RSA Encryption System Using Ancient Indian Vedic Mathematics

Himanshu Thapliyal and M.B Srinivas
(thapliyalhimanshu@yahoo.com, srinivas@iiit.net)
Center for VLSI and Embedded System Technologies
International Institute of Information Technology
Hyderabad-500019, India

## ABSTRACT

This paper proposes the hardware implementation of RSA encryption/decryption algorithm using the algorithms of Ancient Indian Vedic Mathematics that have been modified to improve performance. The recently proposed hierarchical overlay multiplier architecture is used in the RSA circuitry for multiplication operation. The most significant aspect of the paper is the development of a division architecture based on Straight Division algorithm of Ancient Indian Vedic Mathematics and embedding it in RSA encryption/decryption circuitry for improved efficiency. The coding is done in Verilog HDL and the FPGA synthesis is done using Xilinx Spartan library. The results show that RSA circuitry implemented using Vedic division and multiplication is efficient in terms of area/speed compared to its implementation using conventional multiplication and division architectures.

## 1. INTRODUCTION

The standard techniques for providing privacy and security in data networks include encryption/decryption algorithms such as Advanced Encryption System (AES) (private-key) and RSA (public-key) [1,2,3]. RSA is one of the safest standard algorithms, based on public-key, for providing security in networks. While hardware implementation of this algorithm tends to be faster compared to its software counterpart, there is a scope for further improvement of performance of RSA hardware. One of the most time consuming processes in RSA encryption/decryption algorithm is the computation of $a^b$ mod n where a is the text, (b,n) is the key[4] and this paper examines how this computation could be speeded up drawing up on the Indian Vedic Mathematics.

### 1.1 Algorithm to compute $a^b$ mod n efficiently:

One can make use of the following property of modular arithmetic to calculate $a^b$ mod n :

(a1 mod n)*(b1 mod n)] mod n = (a1*b1) mod n

This reduces the intermediate results to modulo n and makes the calculation practical. If we express 'b' as a binary number $b_k b_{k-1} \ldots \ldots b_0$, then we have

$$b = (\sum_{b_j \neq 0} 2^j) \quad \text{Therefore,} \quad a^b = a^{(\sum_{b_j \neq 0} 2^j)} = \prod_{b_j \neq 0} a^{2^j}$$

$a^b$ mod n = [$\prod_{b_j \neq 0} a^{2^j}$] mod n = $\prod_{b_j \neq 0} [a^{2^j} \mod n]$ mod n

The algorithm employed to compute $a^b$ mod n can be summarized as follows:

```
l = 0; m = 1
For j = k downto 0
    do l = 2*l
        m = (m*m) mod n
        if b_j = 1 then  l = l+1
```

```
            m = (m*a) mod n
return m
```

Thus, we can see from the aforesaid algorithm that multiplication and division operations are two of the most important operations in computation of $a^b$ mod n and a high performance multiplication and division algorithm/ architecture will considerably improve the speeds of encryption and decryption. Two known methods of multiplication are array and booth multiplication each with its own limitations [5,6]. Another prominent operation in computation of $a^b$ mod n is division operation and there are several well-known methods for implementing integer multiply and divide circuits. These methods employ operational algorithms with components such as shift registers and adder circuits [7].

From the architecture point of view, division circuits are usually much larger than multiplier circuits for an equivalent data word length and division is generally performed through restoring and non-restoring algorithms [8,9]. These conventional methods of performing digital division typically involve subtracting the divider from a reference number, referred to as a current number, and generally require that the divider be added back to the current number after each computation stage. While in the restoring method, the decision of adding back the divider to the current number depends on the result of the subtraction stage, in non-restoring method, the choice between addition and subtraction is made in the next computation stage, that is, after the subtraction stage. Thus, substantial amount of logic and related logic circuitry are required to implement restoring and non-restoring division algorithms. Also, the circuitry typically involves at least one exclusive-OR (XOR) gate for every bit, which is equivalent of five NAND gates [10].

Further, when the integrated circuit is a field programmable gate array (FPGA), large amounts of valuable programming resources are required to implement digital division, thus limiting the size or precision of the data words that can be accommodated. In addition, the numerous data paths required to connect the related logic circuitry within a FPGA result in slow performance and in some cases may cause it to malfunction.

Thus, in order to circumvent the limitations of conventional multiplication and division algorithms, this paper proposes the implementation of Vedic multiplication and division algorithms that result in improved efficiency [11,12]. A faster and novel hierarchical overlay multiplier has earlier been proposed based on Ancient Indian Vedic Mathematics that performs better than the conventional multiplier architectures [6]. While this paper still utilizes the same multiplier in computation of $a^b$ mod n, it also proposes a novel division algorithm and architecture based on Ancient Indian Vedic Mathematics. It is found that computation of $a^b$ mod n achieves a significant improvement with the inclusion of Vedic division algorithm. The results are found to be quite encouraging than the conventional restore and non-restore divisions.

## 2. RSA ALGORITHM

RSA algorithm can generally be further classified into key generation algorithm, encryption algorithm, and decryption algorithm.
The RSA key generation algorithm can be described in the following five steps.
1. Generate two large random and distinct primes P
   and Q
2. Calculate N = P.Q and K = (P – 1)(Q – 1)
3. Choose a random integer J, 1 < J < K, such that
   gcd(J,K) = 1
4. Compute the unique integer I, 1 < I < K, such that JI= 1 (mod K)
5. Public key is (N, J) and private key is (N, I)

While RSA encryption algorithm is described by
L = $M^J$ mod N,
the decryption algorithm is described by
M = $L^I$ mod N,
where L represents the cipher text and M represents the message.

## 3. MULTIPLIER ARCHITECTURE

The Multiplier Architecture is based on the Vertical and Crosswise algorithm of ancient Indian Vedic Mathematics. In the overlay architecture, grouping of 4 bits at a time is done for both multiplier and multiplicand and thereafter vertical and crosswise algorithm is applied to decompose the whole of the multiplication operation into 4x4 multiply modules. The algorithm is explained in Table-1 for 16x16 bit number. After getting the sub-product bits in parallel from the 4x4 multiply modules, we can employ an efficient method of addition to generate the final 32 bit product. This method can be generalized for NXN bit multiplication where N is a multiple of 4 such as 8,12,16,20,24,....4n. Thus instead of implementing the entire multiplication through a single NXN bit multiplier, we can get the same product efficiently by using the proposed overlay architecture. The advantage of this is that the multiply operation of large number of bits can now be performed by using smaller and efficient 4x4 multiplier.

TABLE 1- 16 x 16 bit Vedic multiplier Using Urdhva Tiryakbhyam

CP- Cross Product (Vertically and Crosswise)

```
A=   A15 A14 A13 A12      A11 A10  A9  A8       A7  A6  A5  A4       A3  A2  A1  A0
           X3                    X2                    X1                    X0

B=   B15 B14 B13 B12      B11 B10  B9  B8       B7  B6  B5  B4       B3  B2  B1  B0
           Y3                    Y2                    Y1                    Y0
                       X3   X2   X1   X0    Multiplicand[16 bits]
                       Y3   Y2   Y1   Y0    Multiplier  [16 bits]
     -------------------------------------------------------------------
     J    I    H    G    F    E    D    C
     P7   P6   P5   P4   P3   P2   P1   P0   Product[32 bits]
```
Where X3, X2, X1, X0, Y3, Y2, Y1 and Y0 are each of 4 bits.

PARALLEL COMPUTATION & METHODOLOGY

1. CP     X0     = X0 * Y0 = A
          Y0

2. CP     X1  X0   = X1 * Y0+X0 * Y1= B
          Y1  Y0

3  CP     X2  X1  X0 =  X2 * Y0 +X0 * Y2 +X1 * Y1=C
          Y2  Y1  Y0

4  CP     X3  X2  X1  X0 = X3 * Y0 +X0 * Y3+X2 * Y1 +X1 * Y2=D
          Y3  Y2  Y1  Y0

5  CP     X3  X2  X1   = X3 * Y1+X1 * Y3+X2 * Y2=E
          Y3  Y2  Y1

6  CP     X3  X2   = X3 * Y2+X2 * Y3=F
          Y3  Y2

7  CP     X3    = X3 *  Y3 =G
          Y3

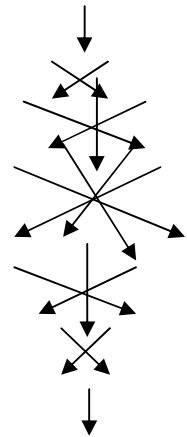

Note: Each Multiplication operation is an embedded parallel 4x4 multiply module

## 4. VEDIC DIVISION ALGORITHM AND ARCHITECTURE

The Division architecture takes N bits of dividend and N bits of divisor to generate the quotient and the reminder. The architecture is based on Straight (At Sight) Division algorithm of Ancient Indian Vedic Mathematics. To simplify the understanding of the algorithm, it is explained in Table-2 for 3 digits by 2 digits number along with an example. The algorithm can be generalized for N digit/bit by N digit/bit number. The processing power of division architecture can easily be increased by increasing the input and output data bus widths since it has a quite a regular structure. Such a regular structure of the architecture has also ramifications for deep submicron designs.

**TABLE 2:   3 digit by 2 digit Vedic Division Algorithm**

X2  X1 X0  by  Y0Y1

```
                    X2        X1:   X0
      Y0                      C1  : C0
      Y1         ________________
                       Z1        Z0  : RD
                 ------------------------------
```

Steps:
1. First do  X2/Y0 (divide)  to get  Z1 as quotient and  C1 as remainder.
2.  Call Procedure ADJUST(Z1,C1,X1,Y1,Y0).
    Now take the next  dividend as
    K=( C1 * 10+X1)-(Y1 * Z1).
3. Do K/Y0(divide) to get Z0 as  quotient and C0 as remainder.
4. Call procedure ADJUST  (Z0,C0,X0,Y1,Y0).
    Now Our required remainder,
    RD=(C0 * 10+X0)-(Y1 * Z1).
Hence the  Quotient= Qt=Z1Z0
          Remainder=RD

Procedure ADJUST (H, I, E, A, B)
{
 While ( (I * 10+E) <  B * H)
  {
    H=H-1;
    I=I+ A;
  }
}

For example   35001/77   will work as follows

```
                3  5   0 0 : 1
        7          7 7 : 7
        7         ---------------
                4   5  4 : 43
```

1.  Divide 35 by 7 and get  5 as the quotient and  0 as the remainder.
2.  Call  ADJUST (5,0, 0,7,7) .
   => modified quotient=5   and  remainder  7
      Next Dividend   K= ( 7 * 10 + 0)-(7 * 4)=42
3.  Do  K/ 7 and get  6 as quotient and  0 as remainder.
4. Call  ADJUST(6,0,0,7,7).
=> modified quotient 5  and remainder 7
Next dividend  K= (7 * 10+0)-(7 * 4)=42
5.  Do  K/7 and get 6 as quotient and 0 as remainder
6. Call  ADJUST (6,0,1,7,7)
=> modified quotient= 4  and remainder  7
 Remainder  RD=(7 * 10+1)-(7 * 4)=43
 Therefore  Quotient =454  and Remainder=43

## 5. VERIFICATION AND IMPLEMENTATION

In this work, while the algorithms have been implemented in Verilog HDL and logic simulation done in Veriwell simulator, synthesis and FPGA implementation have been done using Synopsys FPGA Express. After gate-level synthesis from high level behavioral and/or structural RTL HDL codes, basic schematics have been optimized for speed and area using Xilinx family of devices : SPARTAN, S30VQ100, Speed Grade : -4.  The components of the schematics are FMAP & HMAP which are 4 input and 3 input XOR functions respectively.

## 6.   RESULTS AND DISCUSSIONS

The RSA encryption/decryption circuitry achieves a significant improvement in performance using the Vedic hierarchical overlay multiplier and the novel Vedic division algorithm as reflected by the results shown in Table 3 and Fig. 1. It is found that when implemented with the overlay multiplier architecture and the Vedic division algorithm, the RSA circuitry has less timing delay compared to its implementation using traditional multipliers and division algorithms. It has already been demonstrated that Vedic hierarchical overlay multiplier is efficient than conventional multipliers in terms of area/speed [6]. For the  Xilinx Spartan  family,  it is  found that the gate delay for  RSA circuitry using 8x8 overlay multiplier architecture  and  16 bit by 16 bit Vedic division is  1.507 µs with area of  15235( 14942 FMAP and 293 HMAP). While  the gate delay of  RSA circuitry using  8x8 overlay  multiplier and restore division is  2.838  µs  with area of  14141 (14077 FMAP,64 HMAP) that of non-restore division with 8x8 overlay multiplier is 2.828 µs with area of 6689 (6616, 73 HMAP).

**Table 3: Timing Simulation Results of RSA Circuitry Using Vedic Overlay Multiplier and Division Architectures**

| RSA Architecture (With Overlay Multiplier) | Vendor | Family | Device | Area | | Delay(μs) |
|---|---|---|---|---|---|---|
| | | | | FMAP | HMAP | |
| Restore Division | Xilinx | Spartan | S30VQ100 | 14077 | 164 | 2.838 |
| Non-Restore Division | Xilinx | Spartan | S30VQ100 | 6616 | 73 | 2.828 |
| Vedic Division | Xilinx | Spartan | S30VQ100 | 14942 | 293 | 1.507 |

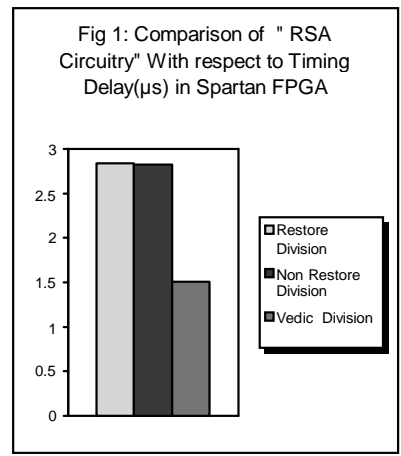

Fig 1: Comparison of " RSA Circuitry" With respect to Timing Delay(μs) in Spartan FPGA

## 7. CONCLUSIONS

The RSA circuitry implemented with Vedic overlay high speed multiplier and novel Vedic division algorithm exhibits improved efficiency in terms of speed and area. Due to its parallel and regular structure the proposed architecture can be easily laid out on silicon chip and can work at high speed without increasing the clock frequency. It has the advantage that as the number of bits increases its gate delay and area increase very slowly as compared to RSA circuitry employing traditional multipliers and division algorithms [6]. It is found that this design is quite efficient in terms of silicon area and speed and should result in substantial savings of resources in hardware when used for crypto and security applications.